\documentclass[aps,amsfonts,prl,showpacs,nobibnotes,nofootinbib,%
tightenlines,twocolumn]{revtex4}
\usepackage{amsmath,amssymb}
\input BoxedEPS
\SetRokickiEPSFSpecial  
\HideDisplacementBoxes
\newcommand{\ten}{\overline{\mathbf{10}}_{f}}
\newcommand{\eight}{\mathbf{8}_{f}}
\newcommand{\Th}{\Theta^{+}}

\newcommand{\half}{\frac{1}{2}}

\newcommand{\beq}{\begin{equation}}
\newcommand{\eeq}{\end{equation}}
\newcommand{\ba}{\begin{array}}
\newcommand{\ea}{\end{array}}

\begin{document}

\title{Diquarks and Exotic Spectroscopy}
\author{Robert Jaffe} 
\author{Frank Wilczek}
 
\affiliation{Center for Theoretical Physics, Laboratory for Nuclear
Science and Department of Physics, Massachusetts Institute of Technology,
Cambridge, Massachusetts 02139\\
MIT-CTP-3401}

\begin{abstract}\noindent We propose that the recently discovered
$\Th$ baryon is a bound state of four quarks and an antiquark,
containing two highly correlated $ud$-pairs.  If so, the $\Th$ has
positive parity, and it lies in an near-ideally mixed $SU(3)_{f}$
$\mathbf{\overline{10}}_{f}\oplus \mathbf{8}_{f}$.  The Roper
resonance and the $P_{11}(1710)$ fit naturally into this
classification.  We predict an isospin $3/2$ multiplet of $\Xi$'s
($S=-2$) with $J^{\Pi}=\half^{+}$ around 1750 MeV. A search for
manifestly exotic $\Xi^{+}$ and $\Xi^{--}$ in this mass range could
provide a sharp test of our proposal.  We predict that charm and 
bottom analogues of the $\Th$ are stable against strong decays.
\end{abstract}

\vspace*{-\bigskipamount}

\maketitle 

An exotic baryon, $\Th (1540)$, with the quantum numbers of $K^{+}n$
has recently been reported by several groups
\cite{Nakano:2003qx,Barmin:2003vv,Stepanyan:2003qr}.  The $\Th$ cannot
be a three quark state.  Its minimal quark content is $uudd\bar s$,
manifestly an exotic.  It has hypercharge two and third component of
isospin zero.  The minimal SU(3)-flavor assignment is at the top of a
$\ten$, also containing a $Y=1$ isodoublet ($N^{+}$, $N^{0}$), a $Y=0$
isotriplet ($\Sigma^{+}$,$\Sigma^{0}$,$\Sigma^{-}$), and a $Y=-1$
isospin-quartet ($\Xi_{\frac{3}{2}}^{+}$, $\Xi_{\frac{3}{2}}^{0}$,
$\Xi_{\frac{3}{2}}^{-}$, $\Xi_{\frac{3}{2}}^{--}$).  In this case the
$\Th$ would be an isosinglet.  The apparent absence of the $I_{3}=+1$,
$\Theta^{++}$ in $K^{+}p$ argues against $I=1$.  We assume the $\Th$
to be an isosinglet.

Exotics with the quark content of the $\Th$ have been proposed in the
context of quark and bag models since the early days of QCD
\cite{Jaffe:1976ii,Strottman:qu}.  Their possible importance in heavy
quark systems (eg.~$suud\bar c$) has been emphasized by Lipkin
\cite{Lipkin:1987sk}.

More recently the SU(3)-flavor $\ten$ representation has emerged as an
interesting feature of chiral soliton (Skyrme) models of baryons. 
Soon after the study of SU(2) chiral solitons by Adkins, Nappi, and
Witten \cite{Adkins:1983ya}, three flavor generalizations led to
predictions of a low lying $J=\half\ \ten$\cite{Chemtob:ar}.  As early
as 1987 Praszalowicz predicted that the $Y=2$ isosinglet member of the
$J=\half\ \ten$ would lie near 1540 MeV \cite{MP}.  In 1997 Diakonov,
Petrov, and Polyakov (DPP) not only predicted a $\Theta$ at about 1540
MeV, but also estimated its width at less than 15
MeV\cite{Diakonov:1997mm,Weigel:1998vt}.

The discovery of a manifestly exotic baryon provides an opportunity to
refine our understanding of quark dynamics at low energy, where it is
not perturbative.  In this Letter we propose that the $\Th$ can be
usefully considered as a bound state of an antiquark with two highly
correlated spin-zero $ud$-diquarks.  This interpretation has some
desirable features and makes predictions that distinguish it clearly
both from the traditional uncorrelated quark model and from the chiral
solition model.

Let us first discuss some general features of the dynamics of a
$K^{+}n$ resonance.  The $\Th$ appears to be remarkably narrow.  It
lies about 100 MeV above $K^{+}n$ threshold at a center of mass
momentum of $k=270$ MeV. Taking a typical range of hadronic
interactions $R$ of 1 fermi, the characteristic parameter $kR$ is
about 1.4.  Assuming isospin zero, the possible quantum numbers of the
$\Th$ (in $KN$ scattering notation, $L_{I(2J)}$) are $S_{01}$,
$P_{01}$, $P_{03}$, $D_{03}$, $D_{05}$, \ldots.  With $kR\approx 1.4$
only the $S$ or $P$ wave is likely, and there is some indication that
the spin of the $\Th$ is $\half$\cite{KHicks}.  At center of mass
momentum 270 MeV the kaon and neutron are not very relativistic. 
There are no other two-body hadronic channels coupling to $K^{+}n$
below $K\Delta$ threshold at 1725 MeV. Since there are no
quark-antiquark annihilation graphs in $K^{+}n$ scattering (that is
what it means to be exotic) one expects, and observes, that low energy
$K^{+}n$ scattering is dominated by ordinary exchange forces.  This
contrasts with non-exotic channels, for example $\pi N$, which are
dominated by confined states that couple by $\bar qq$ annihilation,
for example the $\Delta$.  QCD allows $K^{+}n$ to couple to $q^{4}\bar
q$ configurations with wavefunctions that differ markedly in space,
color, and spin from the $K^{+}n$ scattering state.  Without explicit
introduction of such effects, one has no choice but to describe
$K^{+}n$ scattering in the region of the $\Th$ in terms of
(non-relativistic) potential scattering.  Let us see where that leads. 
Attractive forces in the $s$-wave do not generate resonances.  If they
are attractive enough they produce bound states like the deuteron (or
virtual states like the dineutron), if not, they modulate the phase of
the scattering wavefunction in a smooth fashion.  In higher partial
waves attractive forces typically generate resonances through
interplay between the attraction and the repulsive angular momentum
barrier.  The mass and width of the resonance are related through the
range and depth of the potential.  For a simple attractive potential
(we used a square well) of range 1 fermi, the width of a $P$-wave
resonance 100 MeV above threshold is above 175 MeV. To produce a width
of order 10 MeV requires a range of about $0.05$ fermi, which brings
in a high energy scale.  Thus the observed width of the $\Th$ suggests
that its coupling to the $K^{+}n$ channel is suppressed by some
additional dynamics beyond non-relativistic potential scattering.

Now let us discuss the predictions and reliability of the chiral
soliton and the traditional (uncorrelated) quark models.  In the
chiral soliton model the $(\ten,\half^{+})$ is the third baryon
SU(3)-flavor multiplet in the same ``rotational band'' as the
$(\mathbf{8}_{f}, \frac{1}{2}^{+})$ and $(\mathbf{10}_{f},
\frac{3}{2}^{+})$.  Like them it has positive parity.  The successful
prediction of a narrow $\half^{+}\, \Theta$ near 1540 MeV is
remarkable, but there are reasons for reserve.  At best this model
approximates QCD in the $N_{c}\to\infty$ limit; in addition, it
involves a poorly controlled truncation of the chiral lagrangian,
adiabatic quantization of a system without separation of scales and --
what is especially pertinent -- in the three-flavor case, heavy
reliance on chiral $SU(3)\times SU(3)$ symmetry, which is badly broken
in Nature.  It abounds with states that are exotic for $N_{c}=3$ and,
since they are not observed in Nature, must be discarded as artifacts. 
Even the two flavor model predicts exotic baryons with $(I,J)=
(\frac{5}{2},\frac{5}{2}), (\frac{7}{2},\frac{7}{2})\ldots$.  Difficulties 
that arise in $SU(3)$ chiral soliton models are discussed in 
Refs.~\cite{Itzhaki:2003nr}

The concepts of the quark model are more
obviously grounded in microscopic QCD, and of course this model has
had much success in describing the meson and baryon spectrum.  The
light quark mesons and baryons by and large fall neatly into the
SU(3)-flavor multiplets predicted by the quark model.  In particular,
the ground state multiplets of the $\bar qq$ and $q^{3}$ have the
quantum numbers, including parity, expected of a collection of quarks
(and antiquarks) in the \emph{same eigenstate} of some mean field. 
Thus the lightest mesons -- the pseudoscalar and vector nonets -- have
negative parity; and the lightest baryons (the
($\mathbf{8}_{f},\half$) and the ($\mathbf{10}_{f}, \frac{3}{2}$))
have positive parity.

The nonet of light scalar ($J^{PC}=0^{++}$) mesons with masses below
1000 MeV ($f_{0}(600)$, $f_{0}(980)$, $a_{0}(980)$, $\kappa(800?)$),
however, does not seem to fit a $\bar q q$ description
\cite{Close:2002zu}.  Instead, a heavier scalar nonet above 1000 MeV
seems to be largely $\bar q q$.  The light scalars seem at least to
have a significant component of $\bar q \bar q q q$
\cite{Close:2002zu} dominated by a configuration where the pairs of
quarks (and antiquarks) are separately correlated in a flavor, color,
and spin antisymmetric state \cite{Jaffe:1976ig}.  This diquark is a
boson with color $\mathbf{\overline{3}}_{c}$, flavor
$\mathbf{\overline{3}}_{f}$, and spin zero.  A diquark and antidiquark
neatly couple together to give a nonet of color singlet scalar mesons. 
The apparent fact that the $\bar q \bar q q q$ scalars are lighter
than the $\bar q q$ scalars testifies to the strength of the diquark
correlation.  There is also evidence for strong diquark correlations
in the baryon spectrum \cite{Anselmino:1992vg}.

Attraction between quarks in the color $\mathbf{\overline{3}}_{c}$
channel has profound roots in microscopic QCD. Indeed, by bringing
quarks together in this channel one halves the magnitude of their
effective charge, and thus largely cancels the associated field
energy.  This idea is the basis color superconductivity in dense quark
matter, a context in which it can be made quite rigorous
\cite{Rajagopal:2000wf}.  

With these considerations in mind let us now consider the $q^{4}\bar
q$ system.  In an uncorrelated quark model, in which all the quarks
are in the ground state of a mean field, the ground state of
$q^{4}\bar q$ has \emph{negative} parity.  This is a striking
difference from the chiral soliton model, and from the correlated
quark model we are advocating.  The full $SU(3)_{f}\times SU(2)_{\rm
spin}$ content of this multiplet and bag model estimates of masses and
decay couplings can be found in Ref.~\cite{Strottman:qu}.  There are
very many states in flavor $\mathbf{1}_{f}$, $\eight$, $\ten$,
$\mathbf{27}_{f}$, {\it etc.\/}.  The lightest have the quark content
$uudd\bar d$ and $uudd\bar u$ and would be expected to lie below the
$\Th$.  There is no evidence for a $\half^{-}$ nucleon in this well
explored region of the non-strange baryon spectrum.  Furthermore, all
the light negative parity baryons are well described as orbital
excitations of $q^{3}$.

We propose instead the following picture of the $\Th$ and other
$q^{4}\bar q$ baryons.  The four quarks are bound into two spin zero,
color and flavor $\mathbf{\overline{3}}$ diquarks.  For our purposes
we can regard the diquarks as composite bosons, $[ud]$, $[ds]$, and
$[su]$.  The diquarks obey Bose statistics, but may be expected to
experience a repulsive ``Pauli blocking'' interaction at short
distances.  In order to combine with the antiquark into a singlet, the
two diquarks must combine into a color $\mathbf{3}$ .  Since the
$\mathbf{3}$ is the antisymmetric part of $\mathbf{\overline{3}}
\times\mathbf{\overline{3}}$, the diquark-diquark wavefunction must be
antisymmetric with respect to the rest of its labels.  For
\emph{identical} diquarks, like $[ud]^{2}$ only space labels remain,
and we conclude that the lightest state must have a wavefunction
antisymmetric under space exchange, {\it i.e.\/} with negative
space-parity.  When combined with the antiquark, the resulting
$q^{4}\bar q$ state therefore has \emph{positive} parity.  The three
unlike diquark pairs, $[ud][ds]$, $[ds][su]$, and $[su][ud]$, can be
antisymmetrized in flavor, and therefore can have symmetric, {\it
i.e.\/} positive parity, spatial wavefunctions.  However it is not
unreasonable to expect that the blocking repulsion raises the energy
of the spatially symmetric states.  Evidence, albeit indirect, for
Pauli blocking in the $s$-wave may also come from the apparent absence
of a tightly bound $H$-dibaryon ($[ud][ds][su]$).  We hypothesize that
this effect is strong enough to elevate the resulting negative parity
nonet of $q^{4}\bar q$ states to masses high enough that their effects
on the meson-baryon $S$-wave, where they lie, are weak.  In any case
all these negative parity states have conventional quantum numbers
($\eight$ and $\mathbf{1}_{f}$) and contain at least one strange
quark.  Thus the $Y=1, I=\half$ ``nucleon'' has the quark content
$uuds\bar s$ and would be heavy and weakly coupled to $\pi N$.  We put
these states aside.

\begin{figure}
\begin{center}
\BoxedEPSF{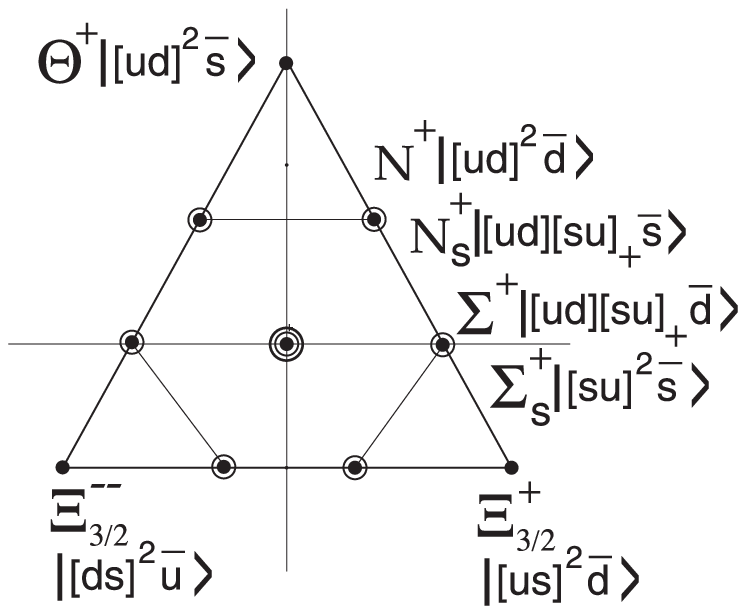 scaled 1050}
\caption{Quark content of representative members of the $(q^{4}\bar 
q)\, \ten + \eight$.}
\label{quarks}
\end{center}
\end{figure}

\begin{figure}
\begin{center}
\BoxedEPSF{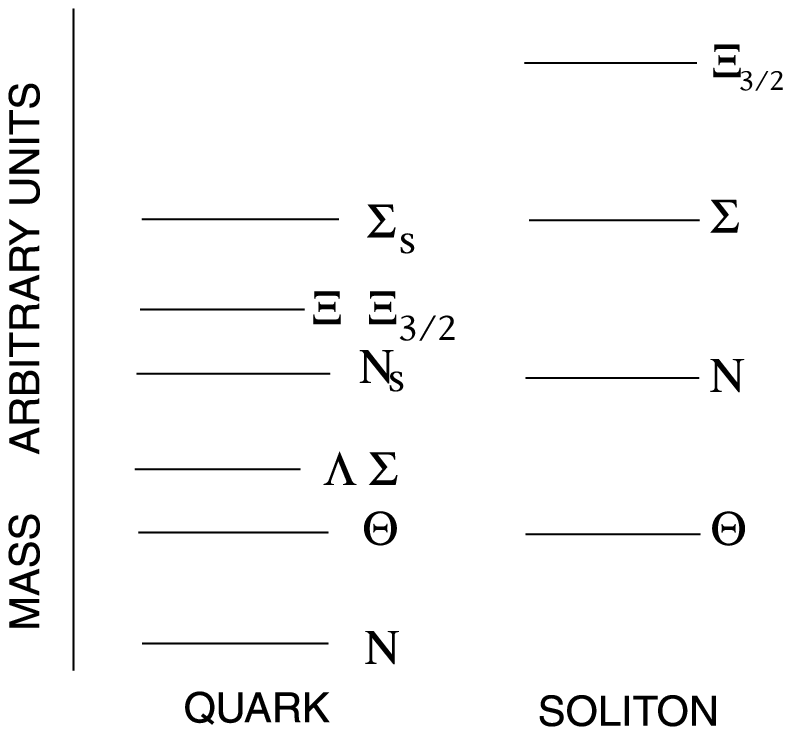 scaled 850}
\caption{Relative masses of states in the ideally mixed $(q^{4}\bar 
q) \ten \oplus \eight$, compared with the mass hierarchy in the chiral 
soliton model.}
\label{mass}
\end{center}
\end{figure}

The flavor symmetric, and therefore spatially antisymmetric, two
diquark states form an $SU(3)_{f}$ $\mathbf{\overline{6}}_{f}$:
$[ud]^{2}$, $[ud][us]_{+}$, $[us]^{2}$, $[us][ds]_{+}$, $[ds]^{2}$,
and $[ds][ud]_{+}$.  When combined with the antiquark the result is a
degenerate $SU(3)_{f}$ $\eight\oplus\ten$.  The strange quark mass
breaks $S(3)_{f}$ symmetry.  In addition to the mass itself, the
$[ud]$ diquark is more tightly bound than the $[us]$ or $[ds]$
diquarks, an effect well-known from baryon
spectroscopy\cite{DeRujula:ge}.  This effect can be related
to the $\Sigma$--$\Lambda$ mass difference.  It leads to a
contribution $\alpha\equiv\frac{3}{4}(M_{\Lambda}-M_{\Sigma})\approx$
60 MeV for every strange quark.  A schematic Hamiltonian including
$SU(3)$ violation is $H_{s}= M_{0}+(n_{s}+n_{\bar s})m_{s}
+n_{s}\alpha$.  It will ideally mix the $\ten$ and $\eight$ according
to their strange quark content.  Of course a more sophisticated 
treatment would allow for non-ideal  mixing and perturb our mass 
estimates.  The quark content of representative
states in the $\ten\oplus\eight$ is given in Fig.~(\ref{quarks}).  The
mass hierarchy of the states is summarized in Fig.~(\ref{mass}) and
compared with the chiral soliton model.  The spatially antisymmetric
wavefunction of the pairs of diquarks would likely correspond to
angular momenum one.  Thus the total angular momentum of the
correlated $(q^{4}\bar q)$ system would be either $\half$ or
$\frac{3}{2}$.  We assume that the $\frac{3}{2}^{+}\,
\eight\oplus\ten$ is elevated to somewhat higher energies, where the
states fall apart quickly into mesons and baryons and do not produce
prominent resonances.  Of course, if our picture proves correct, a
dynamical explanation for this $LS$ splitting must be found.

The $\eight+\ten$ has many interesting features:

$\mathbf{\Theta^{+}([ud]^{2}\bar s)}$: We identify this state with the
$\Th$.  This fixes $M_{0}+m_{s}\approx 1540$ MeV. It may be possible
to explain the narrowness of the $\Th$ by the relatively weak coupling
of the $K^{+}n$ continuum to the $[ud]^{2}\bar s$ state from which it
differs in color, spin and spatial wavefunctions.
    
$\mathbf{N([ud]^{2}\bar d)}$: This is the lightest particle in the
$\eight+\ten$, with mass $M_{0}$.  It has the quantum number of the
nucleon.  It is tempting to identify this state with the otherwise
perplexing Roper resonance, the $N(1440)\ P_{11}$.  The assignment
would explain the existence of the Roper and fixes $M_{0}\approx 1440$
MeV. The $N(1440)$ is much broader than the $\Th$.  The greater phase
space for $N(1440)\to N\pi$ compared to $\Th\to K^{+}n$,
$[p(\Th)/p(N(1440)]^{3}\approx 3$ helps, but does not nearly account
for the width.  Of course, in our picture the internal structure and
group-theoretic properties of $\Th$ and $N(1440)$ are quite different.

$\mathbf{\Sigma([ud][su]_{+}\bar d)}$ and
$\mathbf{\Lambda([ud][ds]\bar d-[su][ud]\bar u)}$: These $\half^{+}$
hyperon resonances should have masses, $M(\Sigma,\Lambda) \approx
M_{0}+m_{s}+\alpha\approx$ 1600 MeV, well above threshold for decay to
$\Lambda(1116)\pi$ and $\Sigma(1192)\pi$.  We expect them to be
broader than the $\Th$.  The PDG identify $\Sigma$ ``bumps'' at 1480
MeV and 1560 MeV. There is a well established $\Lambda(1600)\ P_{01}$ 
and $\Sigma(1660) P_{11}$ both with
$J^{\Pi}=\half^{+}$. 

$\mathbf{N_{s}([ud][su]_{+}\bar s)}$: Although it has the same quantum
numbers as the nucleon, this state has hidden strangeness (like the
$\phi(1020)$) and is therefore heavier and should couple anomalously
strongly to strange particles.   We estimate $M(N_{s})\approx
M_{0}+2m_{s}+\alpha\approx 1700$ MeV. There is a candidate,  the
$N(1710) P_{11}$.  If  this  assignment is correct, then the
$N(1710)$ should couple more strongly to $N\eta$, $K\Lambda$, and
$K\Sigma$ than currently suspected. 

$\mathbf{\Sigma_{s}([su]^{2}\bar s)}$: Like the $N_{s}$, this state
has hidden strangeness.  It is predicted to be heavy, $M(\Sigma_{s})
\approx M_{0}+3m_{s}+2\alpha\approx$ 1850 MeV, and to couple
predominantly to $\Sigma\eta$ and $\Xi\overline K$.  Since
it couples weakly to $\overline KN$, it would have been difficult to see.

$\mathbf{\Xi(\sqrt{2}[us]^{2}\bar u-[su][ds]\bar d)}$: This
$\half^{+}$ cascade resonance should have mass $M(\Xi)\approx
M_{0}+2m_{s}+2\alpha \approx 1750$ MeV. Some $\Xi$'s are known in this
region, but none have established spin-parity assignments.

$\mathbf{\Xi_{\frac{3}{2}}([us]^{2}\bar d)}$: This isospin $3/2$
multiplet contains two $\Xi$'s with ordinary charge assignments,
$\Xi_{\frac{3}{2}}^{0}$ and $\Xi_{\frac{3}{2}}^{-}$.  In addition it
includes the exotic $\Xi_{\frac{3}{2}}^{+}$ and
$\Xi_{\frac{3}{2}}^{--}$.  These exotics should have mass near 1750
MeV. There is a clean $SU(3)$ relation between the matrix elements for
$\Theta^{+}\to pK^{0}/nK^{+}$ and
$\Xi^{--}\to\Sigma^{-}K^{-}/\Xi^{-}\pi^{-}$.  Adjusting for
differences in phase space, the width of the $\Xi^{--}$ at a mass of
1750 MeV should be about 50\% greater than the width of the $\Th$.

The spectrum of the $\eight \oplus\ten$ correlated quark picture differs 
in several dramatic ways from the spectrum predicted by the chiral 
soliton model.
The differences are highlighted in Fig.~(\ref{mass}).  The chiral
soliton model predicts only a $\ten$.  First order perturbation theory
in $m_{s}$ predicts equal spacing between isomultiplets of decreasing
hypercharge.  None of the phenomena related to ideal mixing of the
$\eight\oplus\ten$ occur.  In particular in the chiral soliton model:
a) There is no $N$ lighter than the $\Th$, therefore no candidate for
the Roper resonance; b) The $N$ that is heavier than the $\Theta$
couples both to strange and non-strange channels.  The $N_{s}$ in the
correlated quark picture should couple predominantly with hidden
strangeness (like the $\phi(1020)$ or $f_{0}(980)$); c) The $\Xi$
states are heaviest -- above 2 GeV. In the correlated quark picture
they are close to the $N_{s}$ because they both contain two
$s$-quarks.  The prediction of light charge-exotic $\Xi$'s is the most
distinctive signature of our model; d) There is no second multiplet of
$\Sigma$'s with hidden strangeness and no $\Lambda$.  The striking
departure of predicted mass splittings in the chiral soliton model
from an approximate additive quark mass formula arises from that
model's powerful embodiment of chiral flavor $SU(3)\times SU(3)$.  

Exotic channels ought to be accessible to numerical investigation
using lattice gauge theory.  The predicted resonances are above
threshold, but not by much, so they should be visible in long-range
correlators.  By comparing different spatial and flavor trial wave
functions, one could isolate the effect of the suggested (diquark)
configurations.  One could also bring the states closer to threshold,
or perhaps even below it, by increasing quark masses, as in an
analogous study of the scalar mesons \cite{Alford:2000mm}.

Our picture suggests thatthe charm and bottom analogs of the $\Th$ may be
stable against strong decays.  The $\Theta^{0}_{c}=[ud]^{2}\bar c$ and
$\Theta^{+}_{b}=[ud]^{2}\bar b$ differ from the $\Th$ by the
replacement $\bar s\to\bar c$ or $\bar b$.  The cost of this can be
estimated by comparing the masses of the $\Lambda(1116)$,
$\Lambda_{c}(2285)$, and $\Lambda_{b}(5624\pm 9)$. 
The $[ud]$ diquark in the $\Lambda$ is coupled to color $\mathbf{\bar
3}$ and spin zero and so provides an environment for the $s$ quark
nearly identical to the environment of the $\bar s$ quark in the $\Th$
(plausible in the limit of tight diquark binding).  Therefore we
estimate $M(\Theta^{0}_{c}) -M(\Theta^{+})
=M(\Lambda_{c})-M(\Lambda)=$ 1170 MeV. This yields $M(\Theta^{0}_{c})=
2710$ MeV, which is 100 MeV below threshold for the strong decay
$\Theta^{0}_{c}\to pD^{-}$.  Likewise, we estimate $M(\Theta^{+}_{b})
-M(\Theta^{+}) =M(\Lambda_{b})- M(\Lambda)= 4510\pm 10$ MeV, and
therefore $M(\Theta^{+}_{b})=6050\pm 10$ MeV, which is $165\pm 10$ MeV
below threshold for the strong decays $\Theta^{+}_{b}\to pB^{0}$ and
$nB^{+}$.  The sequence of binding energies, -100 Mev, 100 MeV, and
165 MeV, for the $\Theta^{+}$, $\Theta^{0}_{c}$, and $\Theta^{+}_{b}$
reflects the non-linear dependence of the pseudoscalar meson ($K$,
$D$, and $B$) masses on the quark masses.  The $\Theta^{0}_{c}$ could
be produced, for example, in $\pi^{-} p \to \Theta^{0}_{c}D^{0}$ or
$\Theta^{0}_{c}D^{+} \pi^{-}$ or even in non-leptonic $B$-decays, such
as $B^{0}\to \Theta^{0}_{c} \bar p \pi^{+}$\cite{Karliner:2003si}.

We have presented a description of $q^{4}\bar q$ baryons based on the
strong color-spin correlation force that drives color
superconductivity and may be responsible for the peculiarities of the
light scalar mesons.  A light positive parity $K^{+}n$ exotic is
natural in this picture.  So is an even lighter $\half^{+}$ nucleon,
which we tentatively identify with the Roper resonance.  Our
description differs in important ways from the chiral soliton
description of the $\Th$, and our prediction of relatively light
charge-exotic cascades should be testable in both numerical and
accelerator experiments.

After submission of this Letter, evidence for an exotic $\Xi^{--}$ was
reported by the NA49 Collaboration\cite{unknown:2003vb}, at a mass
about 100 MeV above our rough estimate.

\subsection{Acknowledgments}
We thank K.~Hicks for correspondence and N.~Samios for conversations. 
One of us (RLJ) would like to thank M.~Praszalowicz for conversations. 
This work is supported in part by the U.S.~Department of Energy
(D.O.E.) under cooperative research agreements~\#DF-FC02-94ER40818.



\end{document}